\documentclass[universe,article,accept,pdftex,oneauthor]{Definitions/mdpi} 
\DeclareRobustCommand{\VAN}[3]{#2}
\let\VANthebibliography\thebibliography
\def\thebibliography{\DeclareRobustCommand{\VAN}[3]{##3}\VANthebibliography}

\usepackage{graphicx}	
\usepackage{amsmath}	
\usepackage{bm}

\newcommand{\geffs}{G_{\rm eff}^\Psi} 
\newcommand{\geffsh}{G_{\rm eff}^{\Psi+\Phi}} 

\newcommand{\Lg}{\mathcal{L}}

\newcommand{\ssout}[1]{}

\def\ga{\mathrel{\raise.3ex\hbox{$>$\kern-.75em\lower1ex\hbox{$\sim$}}}}
\def\la{\mathrel{\raise.3ex\hbox{$<$\kern-.75em\lower1ex\hbox{$\sim$}}}}

\def\be{\begin{equation}}
\def\ee{\end{equation}}
\def\bea{\begin{eqnarray}}
\def\eea{\end{eqnarray}}

\def\E{\rm E}
\def\J{\rm J}

\newcommand{\Om} {{\Omega}}
\newcommand{\Ga} {{\Gamma}}

\firstpage{1} 
\pubvolume{1}
\issuenum{1}
\articlenumber{0}
\pubyear{2025}
\copyrightyear{2025}
\datereceived{26 June 2025} 
\daterevised{3 August 2025} 
\dateaccepted{12 August 2025} 
\datepublished{ } 
\hreflink{https://\linebreak doi.org/} 

\Title{The Effects of the Gravitational Coupling Variation on the Local $H_0$ Estimation}

\TitleCitation{The Effects of the Gravitational Coupling Variation on the Local $H_0$ Estimation}

\Author{{Antonio} 
 Enea Romano}

\AuthorNames{Antonio Enea Romano}

 \AuthorCitation{{Romano,} 
 A.E.}

\address{Instituto de Fisica,Universidad de Antioquia,A.A.1226, Medellin, Colombia\\
ICRANet, Piazza della Repubblica 10, I--65122 Pescara, Italy
} 

\abstract{We study the effects of the time evolution of the matter-gravity coupling on the luminosity distance, showing it  can provide a natural explanation to the apparent  Hubble tension. The gravitational coupling evolution induces a modification of the Friedman equation with respect to the $\Lambda$CDM model, {which we study in both the Einstein and Jordan frame.} 
{We consider a phenomenological parametrization of the low redshift variation of the coupling in a narrow redshift shell,} showing how it can affect the distance of the anchors used to calibrate supernovae (SNe), while higher redshift background observations  are not affected.  This effect is  purely geometrical, and it is not related to any change of the intrinsic  SNe physical properties.
The effects of a time varying  gravity coupling only manifest on sufficiently long time scales, such as in cosmological observations at different redshifts, and if ignored lead to  apparent tensions in the values of cosmological parameters estimated with observations from different epochs of the Universe history. }


\begin{document}
\section{Introduction}

The standard cosmological model  based on assuming general relativity and large scale homogeneity and isotropy has proved quite successful in explaining the Universe we observe. Nevertheless there is some increasing evidence that local~{\citep{Riess:2021jrx}} 
 and high redshift~\citep{Planck:2018vyg} estimations  of Hubble parameter $H_0$ are not consistent, {although recent observations are significantly reducing the difference~\citep{Freedman:2024eph}}. Several solutions to explain this tension have been proposed, {for example in terms of early dark energy~\citep{Jiang:2024tll,Liu:2024yan}}, or~local inhomogeneities~\citep{Romano:2016utn}. See~\citep{DiValentino:2021izs,Aluri:2022hzs,Montani:2023xpd} for a review of the vast literature on the subject. Many efforts have been focused on providing an early Universe explanation for this discrepancy, while in this paper we will consider a local solution of the~tension.

We show that a late time variation of the matter-gravity coupling  can have an important effect on the anchors used to calibrate SNe, and~provide an explanation to the tension.
We derive the modified Friedman equation both in the Jordan {and} Einstein frame, to~clarify the relation between cosmological parameters in the two frames, and~use the Jordan frame formulation for calculating observational effects, since it simplifies the calculation of the luminosity distance.
The effect on SNe distance is negligible, since they are located at higher redshift, while the distance of the anchors is modified w.r.t. a $\Lambda CDM$ model, inducing a difference between the local estimation of the Hubble constant, and~the value obtained from higher redshift observations, which are not affected by the local variation of the gravity coupling. {As an application, we consider  a model with a local variation of the gravitational coupling around $z\approx 0.001$, and~show how it can fit well the local $H_0^{loc}$ estimation~\citep{Riess:2021jrx} and SNe data at the same time, with~a value of the parameter $H_0$ compatible with higher redshift observations~\citep{Planck:2018vyg}}.

\section{Varying Matter-Gravity~Coupling}
{Effective} 
 field theory  is a powerful theoretical approach to study the Universe using very general model independent symmetry principles.
The most general Jordan frame EFT quadratic order action~\citep{Gleyzes:2013ooa} for single-field models can be written schematically as 
\be
\Lg_{\rm}=\sqrt{-g_{\J}}\Big[\Omega^{2} R_{\J} + L^{\rm{DE}}_{\J}+L^{\rm matter}_{\J}(g_{\J})\Big]\,\label{DEJ} \,,
\ee
which in the Einstein frame corresponds to 
\be
\Lg_{\rm }=\sqrt{-g_{\E}}\Big[R_{\E} + L^{\rm DE}_{\E}+L^{\rm matter}_{\E}(\Omega^{-2} g_{\E})\Big]\label{DEE}\,,
\ee
{where $\Omega$ plays the role of  effective Planck mass, $g_{\E}$ and $g_{\J}$ denote the metric respectively in the Einstein and Jordan frame}, and~the two frames are related by the conformal transformation $g_{\E}= \Omega^2 \,g_{\J}$, and~we are using units in which $8\pi G=c=1$ .
{
The term $L^{\rm matter}_{\E}(\Omega^{-2} g_{\E})$ in Equation~(\ref{DEE}) indicates that in the Einstein frame matter is not minimally coupled to the metric, i.e.,~indices in the matter Lagrangian are contracted with $\Omega^{-2} g_{\E}$, not just $g_{\E}$. We will discuss in more details the implications of non-minimal coupling in the next section.
}

Physical {observables} should be invariant under conformal transformations, which are just field redefinitions, but~the components of the energy-momentum tensor are not invariant~\citep{Cote:2019kbg}, and~under a generic transformation $\tilde{g}=\Omega^2 g$ they transform as $\tilde{T}_{\mu}^{\nu}= \Omega^{-4} T_{\mu}^{\nu}$. This implies that the field equations obtained by varying the action with respect to the metric in different frames will have different energy-stress tensors on the r.h.s., and~in particular the Friedman equation obtained assuming a FRW background metric, will be different in the two frames. {In the following we will denote with a subscript $\E$ and $\J$ quantities respectively in the Einstein and Jordan frame.}
{
\section{Equations of Motion in Jordan and Einstein~Frame}
In order to understand the effects of conformal transformations let us consider how the equations of motion of massive particle transform. 
The equation of motion for a massive particle minimally coupled to gravity is obtained from the action 
\be
S=m_0\,\int  g_{\mu\nu} \frac{d x^{\mu}}{d\tau} \frac{d x^{\nu}}{d\tau} d\tau \,,
\ee
where we are denoting with $\tau$ the proper time, and~$x^{\mu}(\tau)$ are the coordinates of the particle as a function of proper time. The~above action is the natural covariant generalization of the classical mechanics kinetic energy, and~its variation is equivalent~\citep{inverno:1992} to that of the  action
\be
S=m_0\,\int \sqrt{ g_{\mu\nu} \frac{d x^{\mu}}{d\tau} \frac{d x^{\nu}}{d\tau}} d\tau =m_0\int ds\,,
\ee
where $ds$ denotes the infinitesimal distance, implying that massive particles move along curves minimizing the distance between space-time points, i.e.,~geodesics.
The variation of the action implies the Lagrange equations, which are the geodesics equations
\be
\ddot{x^{\mu}}+\Ga^{\mu}_{\rho\sigma}x^{\rho}x^{\sigma}=0\,,\label{geod}
\ee
where $\Ga^{\mu}_{\rho\sigma}$ denotes the Christoffel connection coefficients.
After a conformal transformation $\tilde{g}=\Omega^2 g$ the action takes the form
\be
S=m_0\,\int  \Omega^{-2} \tilde{g}_{\mu\nu} \frac{d x^{\mu}}{d\tau} \frac{d x^{\nu}}{d\tau} d\tau \,,
\ee
showing that matter is not minimally coupled to gravity in the new frame,
while the Christoffel coefficients
transform as~\citep{Dabrowski:2008kx}
\bea
\label{connections}
\tilde{\Ga}^{\mu}_{\rho\sigma}
&=&
\Ga^{\mu}_{\rho\sigma} + \frac{1}{\Om}\left( \delta^{\mu}_{\rho} \Om_{,\sigma} +
  \delta^{\mu}_{\sigma} \Om_{,\rho} - g_{\rho\sigma}g^{\mu\alpha}\Om_{,\alpha} \right)\,,
\\
\label{connections1}
\Ga^{\mu}_{\rho\sigma}
&=&
\tilde{\Ga}^{\mu}_{\rho\sigma}- \frac{1}{\Om} \left( \tilde{\delta}^{\mu}_{\rho}
  \Om_{,\sigma} + \tilde{\delta}^{\mu}_{\sigma} \Om_{,\rho} -
\tilde{g}_{\rho\sigma}\tilde{g}^{\mu\alpha}\Om_{,\alpha} \right).
\eea
{Using} 
 the above transformations we can obtain the equation of motion (\ref{geod}) in terms of the metric $\tilde{g}$ 
\be
\ddot{x^{\mu}}+\tilde{\Ga}^{\mu}_{\rho\sigma}x^{\rho}x^{\sigma}=\frac{1}{\Om} \left( \tilde{\delta}^{\mu}_{\rho}
  \Om_{,\sigma} + \tilde{\delta}^{\mu}_{\sigma} \Om_{,\rho} -
\tilde{g}_{\rho\sigma}\tilde{g}^{\mu\alpha}\Om_{,\alpha} \right)x^{\rho}x^{\sigma}\,.\label{geodE}
\ee
{The} right hand side of Equation~(\ref{geodE}) indicates that particles do not follow the geodesics corresponding to the metric $\tilde{g}$, which is sometime interpreted as the effect of a fifth force~\citep{Uzan:2020aig}. In~this paper we will define as Jordan frame the one in which matter is minimally coupled to gravity, in~which particles propagate along geodesics, while in the Einstein frame, which according to the general notation introduced above corresponds to \mbox{$g_{\E}=\tilde{g}=\Omega^2 g=\Omega^2 g_{J}$,} particles do not follow geodesics.
The Einstein and Jordan frame calculations of physical observables must agree, since conformal transformations have no physical effect, and~correspond only to a field redefinition, but~it can be useful, although~not necessary, to~compare the two equivalent formulations to understand the difference with respect to general~relativity.
}


 \section{Jordan Frame Modified Friedman~Equation}
In the Jordan frame the  variation of the action w.r.t. the metric gives the field equation 
\be
\Omega^2 G^{\mu\nu}_{\J}=T^{\mu\nu}_{\J}\,,
\ee
from which we obtain the
modified Friedman equation 
\bea
H_{\J}(z)^2&=&{\Bigg[\frac{\Omega(0)}{\Omega(z)}\Bigg]}^{2}H^2_{\J,0}\Big[\Omega_M(1+z)^3+\Omega_R(1+z)^4+\Omega_{DE}(1+z)^{3(1+w)}\Big]\label{MFJ},
\eea
{where $H_{\J}=\dot{a}_{J}/a_{J}$ denotes the Jordan frame Hubble parameter, and~$a_{J}$ is the Jordan frame scale factor.} 
From the null geodesics equation we get that the comoving distance is given~by
\be
r=\int \frac{da_{\J}}{H_{\J} a_{\J}^2} \,,\label{rJ}
\ee
and from the relation between $a_{\j}$ and the redshift we can compute the luminosity distance in a flat  Universe, giving  the standard formula
\be
D_L(z)=(1+z)\int^z_0\frac{dz'}{H_{\J}(z')}\,\label{DLJ}.
\ee

\section{Einstein Frame Modified Friedman Equation and Conservation~Laws}
{Let} us assume a flat FRW metric
\bea
ds_{\J}^2&=& dt_{\J}^2-a_{\J}(t_{\J})^2\gamma_{ij}dx^j dx^j\,.
\eea
{The} results that follow can be easily generalized to a curved universe, so we will just focus on the flat case.
Assuming no interaction between fluids in the Jordan frame, since matter follows the Jordan frame metric geodesics, the~energy momentum tensor is conserved in the Jordan frame~\citep{inverno:1992}
\bea
\nabla_{\mu}T_{\J}^{\mu\nu}&=&0\,, \\
\dot{\rho}_{\J}+3 \frac{\dot{a_{\J}}}{a_{\J}} (\rho_{\J}+P_{\J})&=&0 \label{ecJ} \,,\label{ec}
\eea
where a dot denotes a derivative w.r.t. the Jordan frame time $t_{\J}$.
For a FRW metric the conformal transformation $g_{\E}=\Omega^2 g_{\J}$ 
corresponds to a scale factor redefinition 
\bea
a_{\E}&=&\Omega\, a_{\J} \,, \label{at}
\eea
{where} 
$a_{\E}$ is the Einstein frame scale factor, while the components of a tensor in the two frames are related~\citep{Cote:2019kbg} by
\bea
T^{\quad \mu}_{E,\nu}&=&\Omega^{-4} T^{\quad \mu}_{J,\nu} \label{Tt}\,, \label{pfc}
\eea
which for a perfect fluid imply
\bea
\rho_{\E}=\Omega^{-4}\rho_{\J} &,& 
P_{\E}=\Omega^{-4}P_{\J}\,. \label{TC}
\eea
{Substituting} Equations~(\ref{at}) and~(\ref{pfc}) in Equation~(\ref{ec}) we obtain
\bea
\dot{\rho}_{\E}+3\frac{\dot{a_{\E}}}{a_{\E}} (\rho_{\E}+P_{\E})+(\rho_{\E}-3P_{\E})\frac{\Omega'}{\Omega}&=&0 \,.
\eea
The modification of the continuity equation is due to the non minimal Einstein frame gravity coupling, and~is  the manifestation of the fifth force~\citep{Uzan:2020aig}, or~equivalently of the universal interaction of the scalar  field with any other~field.

For a perfect fluid minimally coupled to the Jordan frame metric the equation of state $P_{\J}=w\, \rho_{\J}$ and the continuity equation imply the well known relation 
\be
\rho_{\J} \propto a_{\J}^{-3(1+w)} \,.\label{rhoJ}
\ee
{In} the Einstein frame we can obtain a similar relation by rewriting the modified continuity equation   in terms of the scale factor
\bea
\frac{d\rho_{\E}}{da_{\E}}\dot{a}_{\E}+3\frac{\dot{a}_{\E}}{a_{\E}} \rho_{\E}(1+w)+\rho_{\E}(1-3w)\frac{d\Omega}{da_{\E}}\frac{\dot{a}_{\E}}{\Omega}=0 \,,
\eea
which gives the solution
\bea
\rho_{\E}(a_{\E}) &\propto& a_{\E}^{-3(1+w)}\Omega^{3w-1}\,. \label{rhoae}
\eea
\textls[-25]{{Note} that Equation~(\ref{rhoae})  can be also obtained directly by combining {Equations~(\ref{rhoJ}),~(\ref{at}) and~(\ref{TC}).}} 

The redshift is related to the scale factor in the two frames by~\citep{Romano:2023ozy}
\bea
(1+z)=\frac{a_{\J}(0)}{a_{\J}(z)}=\frac{\Omega(z)}{\Omega(0)}\frac{a_{\E}(0)}{a_{\E}(z)} \,, \label{zEJ}
\eea
which substituted in Equation~(\ref{rhoae}) gives
\bea
\rho_{\E}(z)=\rho_E(0) (1+z)^{3(1+w)}\Bigg[\frac{\Omega(z)}{\Omega(0)}\Bigg]^{-4} \,,\label{rhoEz}
\eea
in agreement with Equation~(\ref{Tt}).

In the Einstein frame the metric is
\bea
ds_{\E}^2&=& dt_{\E}^2-a_{\E}(t_{\E})^2\gamma_{ij}dx^j dx^j\,.
\eea
where $dt_{\E}=\Omega\, dt_{\J}$.
The first Friedman equation in the Einstein frame takes the form
\bea
H_{\E}^2&=&\frac{1}{3}\sum_i \rho_{\E,i} \label{FE}
\eea
where the Hubble parameter is defined in the Einstein frame as
\be
H_{\E}=\frac{d a_{\E}}{d t_{\E}}\,,
\ee
and $\rho_{\E,i}$ are the energy densities of the different~fluids.

From Equations~(\ref{rhoEz}) and~(\ref{FE}) we obtain the redshift space equation
\bea
H_{\E}(z)^2&=&{\Bigg[\frac{\Omega(0)}{\Omega(z)}\Bigg]}^{4}H^2_{\E,0}\Big[\Omega_M(1+z)^3+\Omega_R(1+z)^4+ 
\Omega_{DE}(1+z)^{3(1+w)}\Big]\label{MFE} \,,
\eea
where  we have defined in the standard way the dimensionless density parameters 
\bea
\Omega_i=\frac{\rho_{\E,i}(0)}{3 H_{\E,0}^2} \,,
\eea
and factorized the common factor $[\Omega(0)/\Omega(z)]^4$. 
As expected, Equation~(\ref{MFE}) reduces to the standard $\Lambda$CDM form when $\Omega(z)=1$, i.e.,~when matter is minimally coupled to the Einstein frame metric $g_{\E}$, but~ if $\Omega(z)\neq 1$ the cosmological parameters $H_{\E,0}$ and $\Omega_i$  will differ from the $\Lambda$CDM ones. 

Note that the modified Friedman equation in Equation~(\ref{MFE}) could be obtained directly from Equations~(\ref{TC}) and~(\ref{FE}), but~the above derivation based on obtaining $\rho_{\E}(z)$ from the Jordan frame conservation equation is useful to understand the physical origin of the redshift space Friedman equation modification, and~to check and interpret it in terms of conservation laws in different frames.
As previously mentioned, note that the Hubble parameter and the density parameters appearing in the Friedman equation are not the same in the two frames due to the conformal transformations of the energy-stress tensor components given in Equation~(\ref{TC}) and the difference between $a_{\E}$ and $a_{\J}$, while physical observables such as the luminosity distance are conformally invariant\citep{Deruelle:2010ht,Chiba:2013mha,Rondeau:2017xck}.

\textls[-15]{Assuming isotropy, photons propagate along null geodesics defined by $ds_{\E}^2=\Omega^2 ds_{\J}^2=dt^2_{\E}-a_{\E}^2 dr^2=0$, implying $dr=dt_{\E}/a_{\E}$, from~which we obtain the standard flat FRW~formula}
\be
r=\int \frac{da_{\E}}{H_{\E} a_{\E}^2} \,.\label{rE}
\ee

\textls[-15]{From Equation~(\ref{zEJ}) we can see that in the Jordan  $dz=-da_{\J}/a_{\J}^2$, allowing to derive Equation~(\ref{DLJ}), while in the Einstein frame $dz$ also depends on $d\Omega$, making more convenient the calculation of the luminosity distance in the Jordan frame, as~we will do in the following~sections.}


\section{$\Omega \Lambda CDM$ Model}
\begin{figure}[H]
\includegraphics[width=\columnwidth]{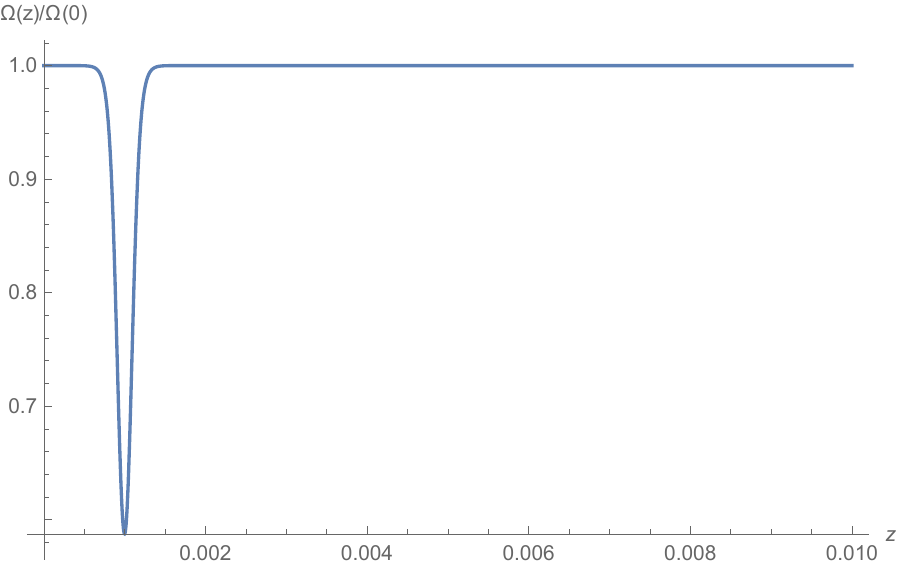}
\caption{The function $\Omega(z)/\Omega(0)$ is plotted as function of redshift. The~gravitational coupling is varying only in a small range of redshift, without~any effect on higher redshift~observations}
\label{fig:Omega}
 \end{figure}

Let's consider a model with a cosmological  constant  $\Lambda_J$ in the Jordan frame, which  gives the modified redshift space  Friedman equation 
\bea
H_{\J}(z)^2={\Bigg[\frac{\Omega(0)}{\Omega(z)}\Bigg]}^{2}H^2_{\J,0}\Big[\Omega_M(1+z)^3+\Omega_{\lambda}\Big]\label{MFELJ}\,.
\eea
{The} corresponding Lagrangian  in the Jordan frame is
\be
\Lg_{\rm }=\sqrt{-g_{\J}}\Big[\Omega^2 R_{\J} -2\,\Lambda_{\J} +L^{\rm matter}_{\J}( g_{\J})\Big]\label{LLJ}\,.
\ee
{{Another} possibility is to define a model with a cosmological constant in the Einstein frame, as~shown in Appendix~\ref{AppA}, and~we leave this case for a future work.}
{The function $\Omega(z)$ is related to the running of the effective Planck mass, and~since local observation such as solar system constraints, or~high redshift observations such as the cosmic microwave background, do not provide strong evidence of a deviation from the Planck mass, we will introduce a transient modification, in~order to satisfy other existing observational constraints.}
{For this reason} we will model the evolution of $\Omega(z)$ with this parametrization  
\be
\Omega(z)^2=\Omega(0)^2 \left\{1+\lambda  \Bigg[\tanh \left(\frac{z-z_0+\Delta z}{\sigma }\right)-\tanh \left(\frac{z-z_0-\Delta z}{\sigma }\right)\Bigg] \right\}\label{Opar}\,,
\ee
corresponding to a local variation around $z_0$, and~an asymptotic value equal to $\Omega(0)$, as~shown in Figure~\ref{fig:Omega}. 
We will call this $\Omega \Lambda CDM$ model.
{Note that the luminosity distance is given by the integral in Equation~(\ref{DLJ}), so it is natural to expect that for object located at $z\gg z_0$ the local variation of $\Omega(z)$ has a small effect on $D_L(z)$, since most of the integral is unaffected, because~Equation~(\ref{HM}) gives the standard $\Lambda$CDM Hubble parameter for most of the integral range.  Only objects located inside the $\Omega(z)$ local variation, i.e the calibrators, are affected by it.}
This is different form the step models which have been studied previously~\citep{Paraskevas:2023aae}. Another difference is that here we study the geometrical effects on the luminosity distance, and~consequently on $H_0$, while in other studies~\citep{Marra:2021fvf} it was considered a sudden transition of a different definition of effective gravitational constant, with~no cosmological consequences, only affecting the physics of SNe, in~particular their absolute~magnitude.

\begin{figure}[H]
\includegraphics[width=\columnwidth]{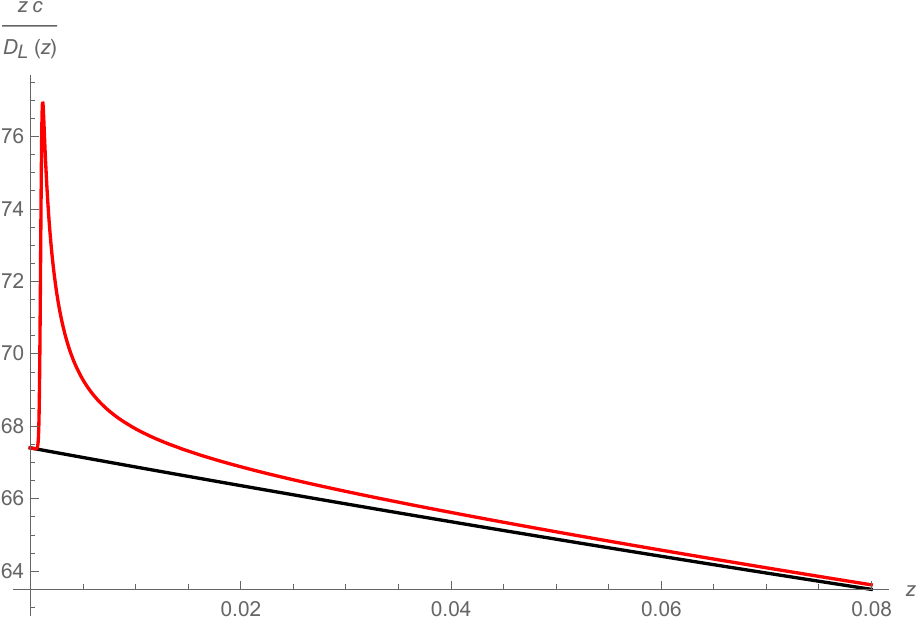}
 \caption{The inverse slope of the luminosity distance is plotted as  as function of redshift for a $\Lambda CDM$ model (black) and and $\Omega \Lambda CDM$ shell model (red), in~units of $H_{\E,0}$.  At~low redshift this is giving the value the Hubble parameter estimated using luminosity distance observations~\citep{Riess:2021jrx}. Inside the shell the value of $H_0^{loc}$  of the $\Omega \Lambda CDM$ shell model is modified w.r.t. the $\Lambda CDM$ model, explaining the Hubble tension, but~at higher redshift  the effect is asymptotically~negligible.}      
 \label{fig:H0loc} 
 \end{figure}

\begin{figure}[H]
        \includegraphics[width=\columnwidth]{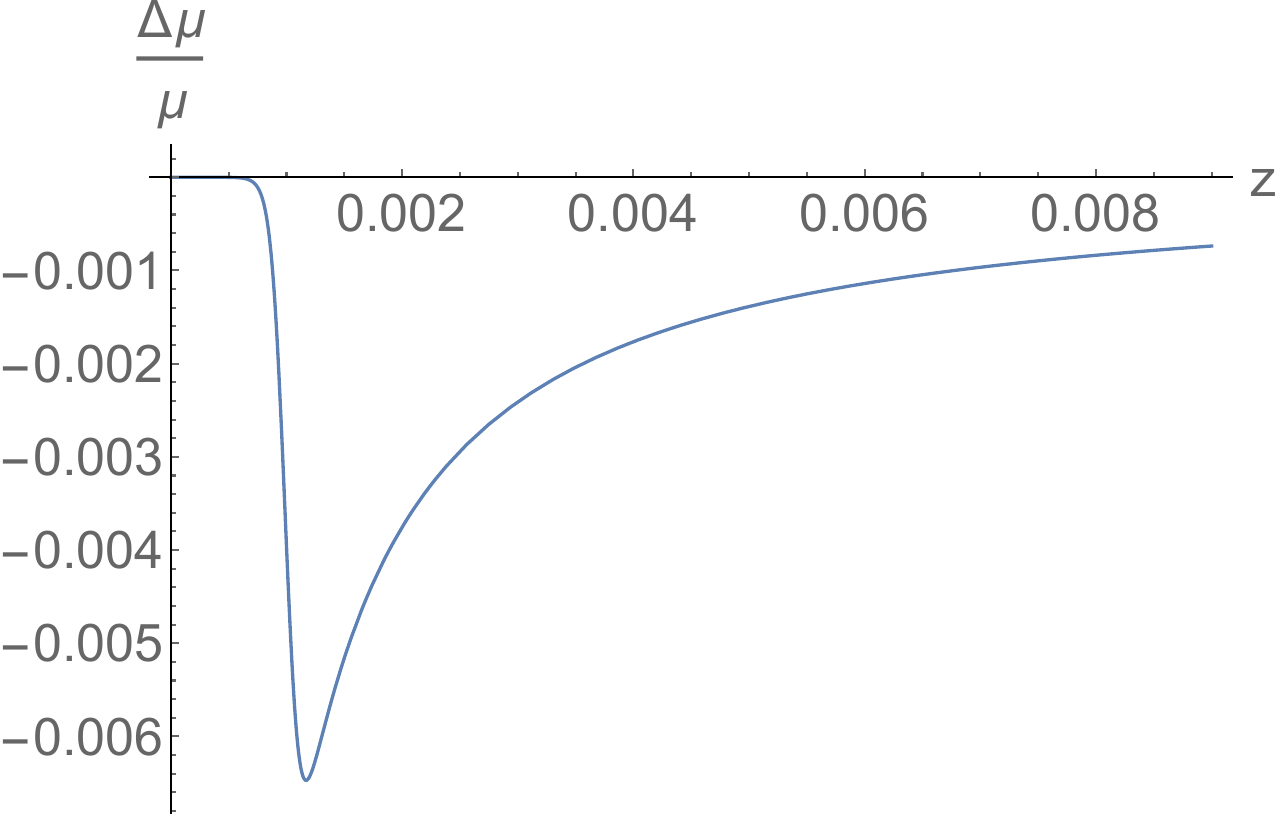}
 \caption{{The relative difference between the  distance modulus $\mu$ of  a $\Omega \Lambda CDM$ shell model and a $\Lambda CDM$ model is plotted as a function of redshift.  The~difference is asymptotically negligible, so only objects inside the shell are affected, i.e.,~anchors such as Cepheids and the megamaser.}}      
 \label{fig:dmu} 
 \end{figure}

\begin{figure}[H]
        \includegraphics[width=\columnwidth]{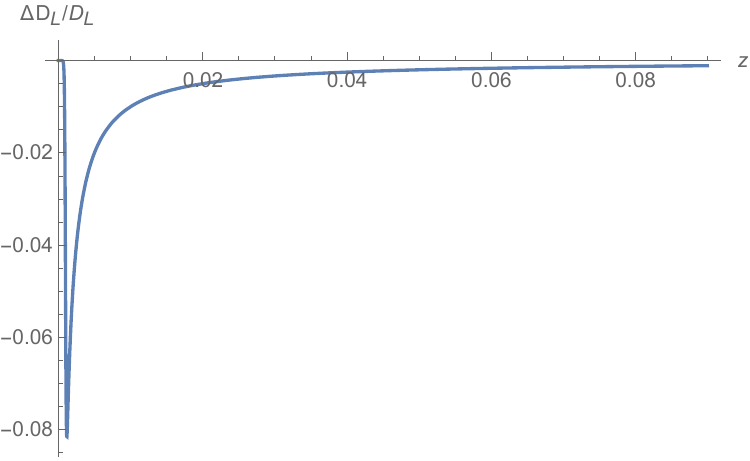}
 \caption{The relative difference between the luminosity distance of   a $\Omega \Lambda CDM$ shell model and a $\Lambda CDM$ model is plotted as a function of redshift.  The~difference is asymptotically negligible, so only objects inside the shell are affected, i.e.,~anchors such as Cepheids and the~megamaser.}      
 \label{fig:DDOD} 
 \end{figure}

The low redshift estimation of the Hubble parameter~\citep{Riess:2021jrx} $H_0^{loc}$, is based on a linear fit of the distance redshift relationship, i.e.,~
\be
H^{loc}(z)=\frac{z \, c}{D_L(z)}\,. \label{H0loc}
\ee
{Note} that this definition corresponds to the actual method to estimate $H_0$ from low redshift SNe distance, and~in general can be different from  $H(z=0)$, which is the quantity considered in other studies of the effects of modified gravity~\citep{Schiavone:2022wvq}.
In  Figures~\ref{fig:Omega} and~\ref{fig:H0loc} we show the plot of the function $\Omega(z)$ and of  $H_0^{loc}(z)$ for the model corresponding to $\lambda = -0.43, z_0 = 0.001, \Delta z = 0.0001$, and~$\sigma = 0.0001$. Inside the shell the value of $H_0^{loc}$  of the $\Omega \Lambda CDM$ shell model is modified w.r.t. $H_{\J,0}$, but~at higher redshift  the effect is asymptotically negligible, as~shown in Figures~\ref{fig:dmu} and~\ref{fig:DDOD}, so the rest of the cosmological parameters $\Omega_i$ are expected {not} to be significantly affected by this kind of $\Omega(z)$ evolution.



\unskip
 
\unskip
 
\unskip
 
\unskip

\section{Effect on SNe~Calibration}
The variation of the gravity coupling at very low redshift is affecting the  distance redshift relation of the anchors used to calibrate SNe, while their distance is not directly affected, because~at higher redshift the distance is the same as in the $\Lambda$CDM mode, as~shown in Figure~\ref{fig:DDOD}. This effect on calibration is propagating on the SNe distance estimation, and~consequently on the estimation of $H_0$. For~a given observed apparent magnitude there is a degeneracy between the absolute luminosity $M$ and $H_0$, i.e.,~the same data  is compatible with different sets of $\{M,H_0\}$
related by~\citep{Mazo:2022auo}
\bea
M_{a} &=& M_{b} + 5\log_{10}\left(\frac{H_{a}}{H_{b}}\right) \label{HM}\,,
\eea
where the subscripts denote the values of different set of~parameters.

This degeneracy is broken by including different observational data sets, such as CMB or calibrating SNe with independent distance anchors. The~Hubble tension is related to the difference between the values of $\{M,H_0\}$ obtained in joint analysis with cosmic microwave background (CMB) data or with low redshift anchors. 
The value of the  parameters corresponding to these different estimations of $H_0$ are reported in Table~\ref{Tpar}. 
\begin{table}[H]
    \caption{{Values} 
 of $\{H_0,M\}$  obtained with different datasets. The~first row shows the values from ~\citep{Riess:2021jrx}, and~the second row the value of $H_0$ from~\citep{Planck:2018vyg} and the implied value of $M$ obtained using Equation~(\ref{HM}). The~ values obtained in previous observational data analysis are underlined, while the value of $M$ for Planck, is inferred using Equation~(\ref{HM}), and~is not~underlined.}
\label{Tpar}    
   \begin{tabularx}{\textwidth}{LCC}
\toprule
\textbf{Dataset} & $\bm{H_0$} \textbf{(km s}$\bm{^{-1}$} \textbf{Mpc}$\bm{^{-1}$}\textbf{)} 
& \textit{\textbf{{M}}} 
 \\
\midrule
Riess & \underline{ $73.04 $ } & \underline{$-19.25 $}\\
Planck & \underline{ $67.4$} & $-19.42$  \\
\bottomrule
\end{tabularx}

\end{table}

As shown in Figure~\ref{fig:H0loc}, the~luminosity distance of anchors is modified w.r.t. to the $\Lambda$CDM value, affecting the local estimation of $H_0$, and~consequently of $M$, because~of Equation~(\ref{HM}). 

\section{Test with SNe~Data}
The $\Omega\Lambda$CDM model is introducing a low redshift modification of the distance redshift relation which could potentially be incompatible with SNe observations. Nevertheless, due to the fact there are no SNe in that redshift range, it is expected that it should not affect significantly the goodness of fit, since the effects on the luminosity distance at higher redshift are negligible, as~shown in Figure~\ref{fig:DDOD}.

\begin{figure}[H]
\includegraphics[width=\columnwidth]{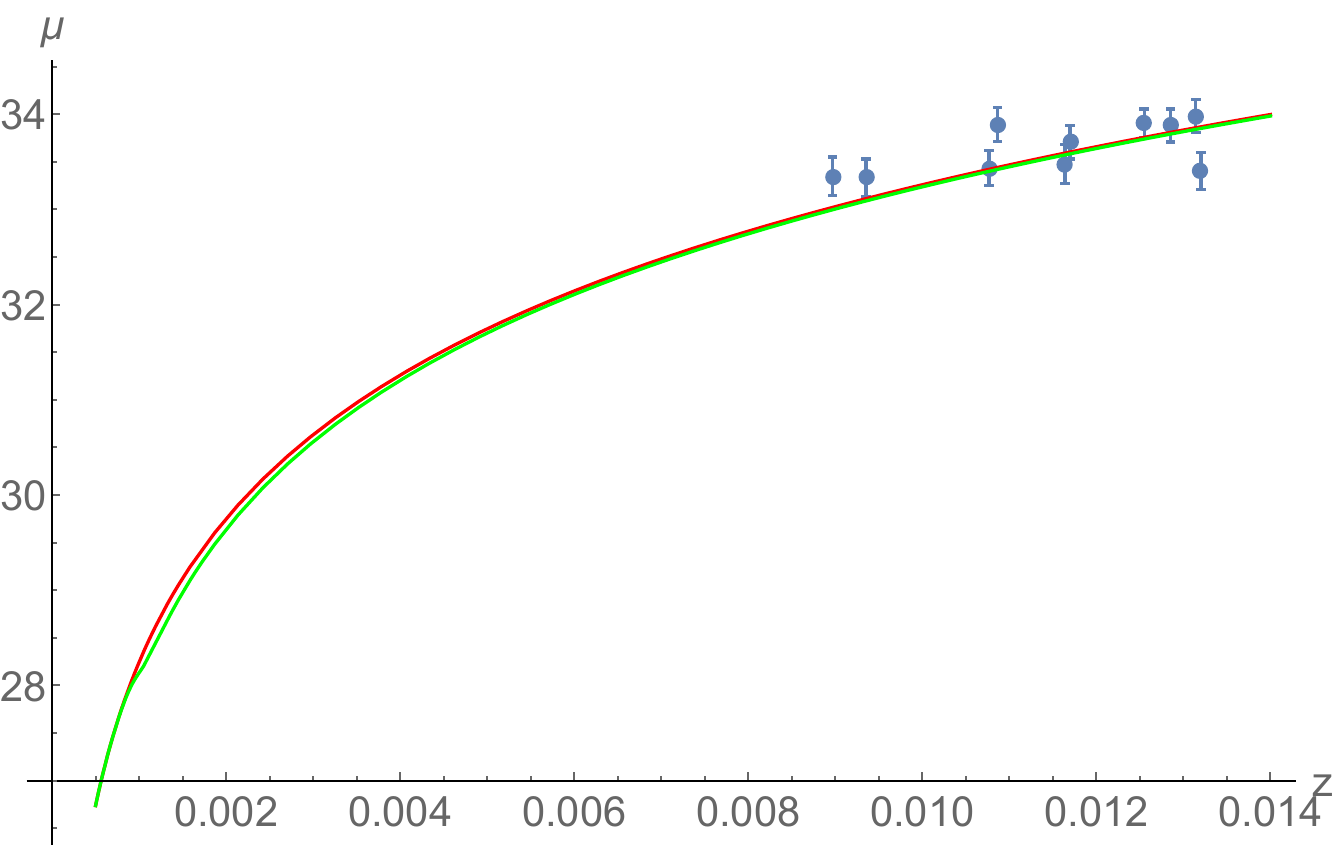}
\caption{{The low redshift  SNe~\citep{Scolnic_2018} distance modulus $\mu$  is compared with different  theoretical models. The~red line corresponds to the $\Lambda$CDM and the green to the $\Omega\Lambda$CDM model, both with Planck parameters corresponding to the second row of Table~\ref{Tpar} . The~two models give very similar predictions for $z>0.009$, so that the only objects affected by the variation of the gravity coupling are those located at $z<0.009$, i.e.,~the anchors, in~agreement with Figure~\ref{fig:DDOD}. The observational data points and their errors are plotted in blue.}
 }

\label{fig:SNlow}

\end{figure}

\begin{figure}[H]
\includegraphics[width=\columnwidth]{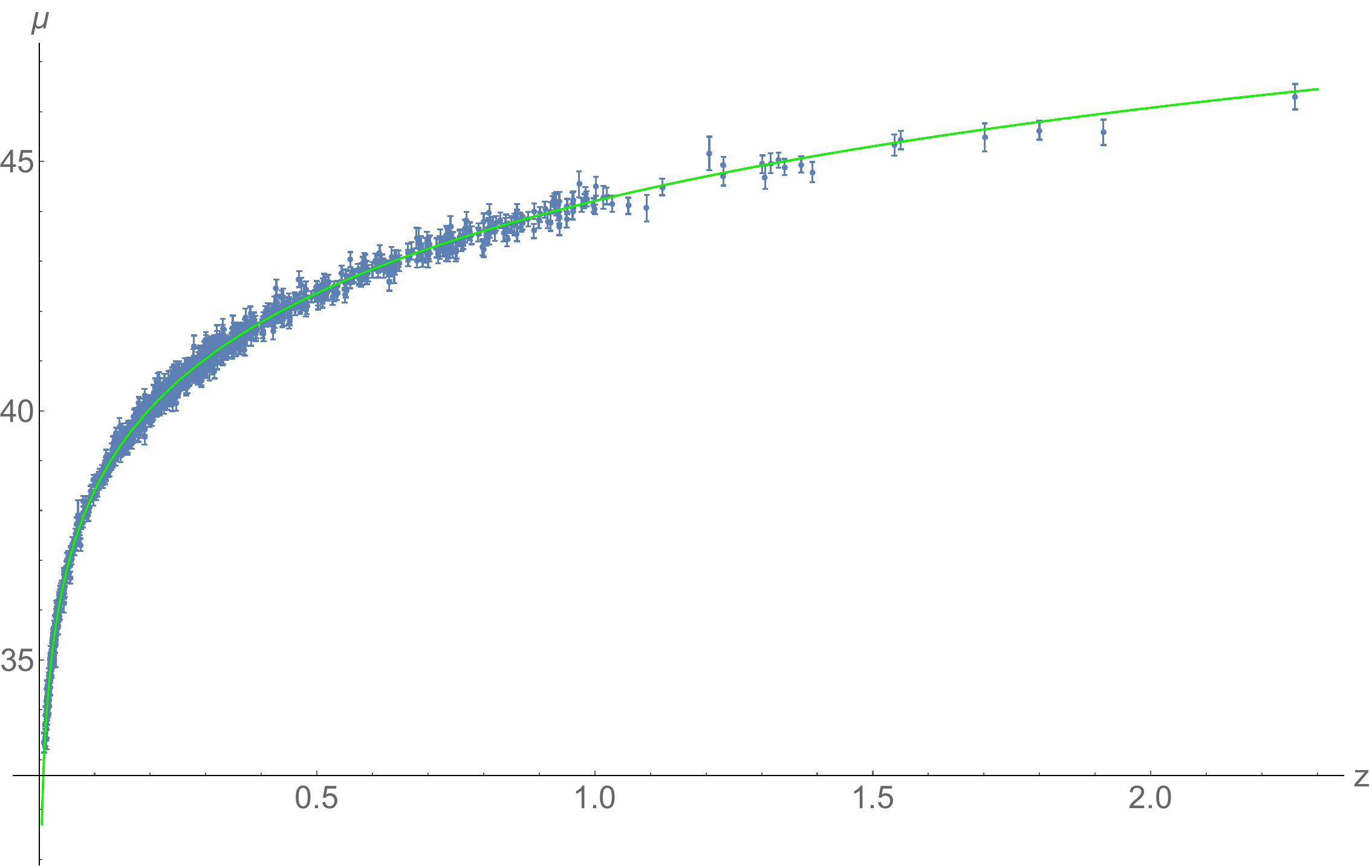}
\caption{ 
{The SNe~\citep{Scolnic_2018} distance modulus $\mu$ is compared with different  theoretical models. {The~red line corresponds} 
 to the $\Lambda$CDM and the green to the $\Omega\Lambda$CDM model, both with Planck~\citep{Planck:2018vyg} cosmological parameters corresponding to the second row of Table~\ref{Tpar}. The~two models give very similar predictions for all SNe, and~the two lines are undistinguishable at the scale of this plot. The~only objects affected by the variation of the gravity coupling are those located at $z<0.009$, i.e.,~the anchors, in~agreement with Figure~\ref{fig:DDOD}. See Figure~\ref{fig:SNlow} for a low redshift plot, where the two curves are distinguishable. The observational data points and their errors are plotted in blue.}}

\label{fig:SNall}

\end{figure}

For this purpose we test the model with the Pantheon dataset~\citep{Scolnic_2018}, computing the  $\chi^2$ according to
\begin{equation}
\chi^2_{SN} =\sum_{i,j} [m_i-m^{th}(z_i)]  C^{-1}_{ij} [m_j-m^{th}(z_j)] \,.
\label{chi2SN} 
\end{equation}
{In} the above equation $C$ is the covariance matrix, $m_i$ and $z_i$ are the observed apparent magnitude and redshift, and~$m^{th}$ is the theoretical  apparent magnitude.
The local value of $H_0$ is fitted with this expression for the $\chi^2$
\be
\chi^2_{H_0}=\left(\frac{H_0^{loc}-H_0^{loc,obs}}{\sigma_{H_0^{loc,obs}}}\right)^2 \,.
\label{chi2H0}
\ee
{We} show the comparison between different models in Table~\ref{tab:chi2}. We fix the cosmological parameters to the values obtained by analyzing the Planck mission data~\citep{Planck:2018vyg}, except~for the value of $H_0$, which we vary to compare different models. We leave to a future work the full analysis of different cosmological observations, but~as discussed in the next section, higher redshift observations are expected to be negligibly affected by the low redshift variation of $\Omega(z)$, so that the check of the compatibility of $SNe$ data is the most important one. {The results of the SNe data analysis are shown in Figures~\ref{fig:SNlow} and~\ref{fig:SNall}.}

\unskip

\unskip

\begin{table}[H]
\caption{The $\chi^2$ for different models is reported for SNe and $H_0^{loc}$, {with $\chi_{SN}^2$ and $\chi_{H_0}^2$ defined respectively in Equations~(\ref{chi2SN}) and~(\ref{chi2H0}). We define $\chi_{Tot}^2=\chi_{SN}^2+\chi_{H_0}^2$ and denote with $\chi_{red}^2$ the reduced $\chi_{Tot}^2$}. The~value of $H_0^{loc}$ is obtained evaluating Equation~(\ref{H0loc}) at  = 0.001, corresponding to the anchors used to calibrate SNe. Note that $\Lambda$CDM models with different sets of $\{H_0,M\}$, given  in Table~\ref{Tpar}, have the same $\chi^2_{SN}$ because of the degeneracy given in Equation~(\ref{HM}). The~difference of the total $\chi^2$ between the first and second row is a manifestation of the Hubble tension,  while the third row shows that a $\Omega\Lambda$CDM can fit well the value of $H_0^{loc}$ obtained in~\citep{Riess:2021jrx} with a value of $H_{{\J},0}$ compatible with CMB observations~\citep{Planck:2018vyg}, resolving the~tension.}    
\label{tab:chi2}
\begin{tabularx}{\textwidth}{LCCCCCC}
\toprule
\textbf{Model} & $\bm{H_{\J,0}$}  & $\bm{H_0^{loc}$} & $\bm{\chi_{SN}^2$} & $\bm{\chi_{H_0}^2$} & $\bm{\chi^2_{Tot}$}  & $\bm{\chi^2_{red}$} \\
\midrule
$\Lambda$CDM  & 73.04 & 73.04 &1073.6  & 0    & 1073.6 & 1.0264\\ 
$\Lambda$CDM & 67.4 & 67.4 & 1073.6   & 29.9 & 1103.5 & 1.055\\
\midrule
$\Omega\Lambda$CDM & 67.4 & 72.9 & 1070.8  & 0.02& 1070.82 & 1.0257\\ 
\bottomrule

\end{tabularx}

\end{table}
\unskip

\section{Compatibility with Other~Observations}
The variation of the gravity coupling we have studied is affecting a  narrow low redshift range, as~shown in Figures~\ref{fig:Omega} and~\ref{fig:DDOD}, so that early Universe observations such as Big Bang Nucleosynthesis (BBN) and Cosmic Microwave Background (CMB) are not affected by~it. 

In fact at high redshift $\Omega(z)=\Omega(0)$, so that the modified Friedman equation reduces to the $\Lambda$CDM Friedman equation. Since at high redshift the luminosity distance is the same of a  $\Lambda$CDM model, as~shown in Figure~\ref{fig:DDOD}, the~distance to the last scattering surface is not affected by the low redshift variation of $\Omega(z)$, and~the fit of CMB data should be very closed to that of a $\Lambda$CDM model with the same cosmological parameters.
In regard to BBN, in~the early Universe $\Omega(z)=\Omega(0)$, so that the late time variation of $\Omega(z)$ has no effect on the early Universe formation of~nuclei.

Other studies~\citep{Marra:2021fvf} have shown that a variation of the effective gravitational constant can alleviate the growth tension, and~in the future it will be important to investigate the effects of a local variation of $\Omega$ on cosmological perturbations, in~particular on the matter power spectrum, but~due to its narrow redshift range these effects are not expected to be very~important.

{
\section{Large Scale Structure Constraints on the Effective Gravitational~Coupling}
The function $\Omega(z)$ is related to the effective gravitational constant measured by large scale structure (LSS), and~can be computed using the EFT of dark energy. These observations can hence provide important independent constraints on $\Omega(z)$, to~check if the parameters solving the Hubble tension are compatible with  LSS.
}

The modification of gravity induces a  modification of the  Poisson's equations  ~\citep{Linder:2015rcz}
{
\bea 
& &   \nabla^{2} \Psi = 4\pi a^{2} G^{\Psi}_{\rm eff} \rho_m\,\delta_ m \,, \\  
& &   \nabla^{2} \Phi = 4\pi a^{2} G^{\Phi}_{\rm eff} \rho_m\,\delta_ m \,, \\
& &   \nabla^{2} (\Psi+\Phi) = 8\pi a^{2} G^{\Psi+\Phi}_{\rm eff} 
\rho_m\,\delta_m \,, 
\eea 
with the effective gravitational constant given by~\citep{Linder:2015rcz}
\be 
\frac{G^{\Phi}_{\rm eff}}{G_N}= \frac{2M_p^2}{M_\star^2} 
\frac{[\alpha_B(1+\alpha_T)+2(\alpha_M-\alpha_T)]+\alpha_B'}{(2-\alpha_B)[\alpha_B(1+\alpha_T)+2(\alpha_M-\alpha_T)]+2\alpha_B'} \,. \label{eq:geff}
\ee 
{The} theories defined in Equation~(\ref{DEJ}) correspond to $\alpha_T=\alpha_B=0$,  giving
\be
\frac{G^{\Psi}_{\rm eff}}{G_N}=\frac{2-\bar{\eta}}{\bar{\eta}}\frac{1}{\Omega^2}
\ee
where 
\be 
\bar\eta=\frac{2\Psi}{\Psi+\Phi}=\frac{\geffs}{\geffsh} \label{bareta}
\,.
\ee
{The} quantities $\geffs$ and $\geffsh$ can be constrained by LSS and lensing observations respectively~\citep{Ishak:2024jhs}, but~the parameterizations adopted therein are designed to study deviations from general relativity on cosmological scales, and~are hence quite different from the local narrow variation in Equation~(\ref{Opar}). It is hence difficult to compare those constraints to the values of the parameters which can solve the Hubble tension using the parametrization in Equation~(\ref{Opar}), and~a new analysis is required.
Since the variation of $\geffs$ is actively investigated, it is still important to study its effects on the luminosity distance, and~hence on $H_0^{loc}$, even if these effects may not be large enough to solve the Hubble tension, because~of LSS constraints on $\geffs$.
}

{
For the purpose of testing the compatibility of LSS data with  a local variation of the type we have shown to be able to solve the Hubble tension, a~redshift bins analysis  of LSS and lensing observations would be required, with~particular attention to data around $z\approx0.001$, and~we leave this to a future work.  The~scope of this paper is indeed to investigate the effects of the variation of the gravitational coupling on the luminosity distance, and~determine what form of $\Omega(z)$ would be necessary to solve the Hubble tension, while at the same time fitting well SNe and other background observations such as the distance to the last scattering surface measured by the CMB, which is unaffected, as~shown in Figure~\ref{fig:DDOD}.
}
\section{Implications for the Apparent Hubble~Tension}

The effect of the $\Omega$ shell is to change  $H^{loc}$ w.r.t. $H_{\J,0}$, while  asymptotically the luminosity distance is unaffected, and~consequently high redshift observations such as the CMB will give a value of the Hubble parameter equal to $H_{\J,0}$.
In a $\Lambda$CDM model at low redshift $H^{loc}\approx H_{\J,0}$, and~the well known tension arises.
A small time variation of $\Omega$  explains naturally the apparent Hubble tension within the framework of the $\Omega\Lambda$CDM model.
Ignoring the redshift dependence of $\Omega(z)$ and fitting observational data with the $\Lambda CDM$ model can lead to the apparent discrepancy between low and high redshift estimations of $H_0$. 

Note that the local estimation of $H_0$ is crucially dependent on geometrical distance anchors~\citep{Efstathiou:2021ocp}, such as the megamaser NGC 4258, which are located at a redshift $z_{an}\approx 0.001$. This implies that the Hubble tension can be resolved by a $\Omega\Lambda$CDM model with parameters values such that the shell includes the anchors, i.e.,~for example  $z_0 \approx z_{an}$.


\section{Conclusions}
We have shown that the time variation of the gravity coupling can provide a natural explanation to the apparent tension between the values of cosmological parameters estimated from observations corresponding to different epochs of the Universe history.
We have given an example of a $\Omega$ shell model, {with strong variation of the matter-gravity coupling in a very narrow redshift range centered at $z\approx  0.001$}, which can explain the difference between the local estimation of $H_0$ based on luminosity distance observations, and~high redshift estimations, due to the effects on the SNe  distance anchors.
Since the variation of the gravity coupling {is assumed to occur} only at very low redshift, high redshift observations such as BBN and CMB are not affected by it. The~model can fit well SNe data, since they are located at higher redshift, so that the variation of $\Omega$ has an appeciable effect only on the distance anchors used to calibrate SNe, and~consequently on the value of $H_0^{loc}$.
While the local variation of $\Omega$ is expected to have only negligible effects on high redshift observations, the~full analysis of all available observational data sets is important to confirm the results obtained in this paper analyzing SNe data. We leave this task to a future upcoming~work.

While in this paper we have focused on the effects on the background evolution, in~order to estimate the effects on other cosmological observables, it will also be necessary to compute the effects on the evolution of cosmological perturbations, and~to investigate its effects on the growth tension~\citep{Marra:2021fvf}.
In this paper, inspired by the EFT, we have adopted  a phenomenological approach in modeling the observational effects of $\Omega(z)$, but~in the future it will be important to investigate the fundamental origin of its  variation, considering specific modified gravity~theories.

\vspace{6pt}

\funding{{This work was supported by the UDEA  project 2023-63330.}} 

\dataavailability{The SNe data analyzed in this paper are publicly available at \url{https://github.com/dscolnic/Pantheon} {(accessed on 15 July~2024).}} 

\acknowledgments{I thank the Osaka University Theoretical Astrophysics Group and the Yukawa Institute for Theoretical physics for their kind hospitality. 
I thank Theodore Tomaras and for interesting discussions about the difference between theories with a cosmological constant in  Einstein or Jordan frame, and~Mairi Sakellariadou for the suggestion to compare to observational~data. }

\conflictsofinterest{{The authors declare no conflicts of interest. The funders had no role in the design of the study; in the collection, analyses, or~interpretation of data; in the writing of the manuscript; or in the decision to publish the~results.}} 


\appendixtitles{yes} 
\appendixstart

\appendix
\section{Einstein Frame Cosmological Constant~Model}
\label{AppA}
{Alternatively we could also consider the case of an Einstein frame cosmological constant given by
\be
\Lg_{\rm }=\sqrt{g_{\E}}\Big[R_{\E} -2\Lambda_E +L^{\rm matter}_{\E}(\Omega^{-2} g_{\E})\Big]\label{LLE}\,,
\ee
which corresponds to the special case of Equation~(\ref{DEE}) in which $L^{\rm DE}_{\E}=-2\Lambda_E$.
}
{
In this case the non minimal coupling only affects the matter part of the energy-momentum tensor, not the dark energy part, since the cosmological constant term  is the same as in general relativity, and~the modified Friedman equation takes the form
\bea
H_{\E}(z)^2=H^2_{\E,0}\Bigg[{\Bigg(\frac{\Omega(0)}{\Omega(z)}\Bigg)}^{4}\Omega_M(1+z)^3+\Omega_{\lambda}\Bigg]\label{MFELE}\,.
\eea
{Note} that the dark energy term  in Equation~(\ref{MFELE}) apparently differs from the $w=-1$ limit of Equation~(\ref{MFE}), but~the two equations are actually consistent.
Accounting for the metric determinant transformation  $\sqrt{-\tilde{g}}=\Omega^{-4}\sqrt{-g}$ 
under a conformal transformation $\tilde{g}=\Omega^2 g$, the~Einstein frame cosmological constant 
Lagrangian $\sqrt{-g_{\E}}\Lambda_E$ corresponds to $\sqrt{-g_{\J}}\Omega^{4}\Lambda_E$ in the Jordan frame (since $g_{\J}=\Omega^{-2} g_{\E}$), i.e.\, $\rho_J^{\Lambda_E}\propto \Omega^4$ is not a constant, so that Equation~(\ref{rhoEz})~gives
\be
\rho^{\Lambda_E}_E(z) \propto (1+z)^{3(1+w)}\Omega^{-4}\Omega^{4}\,,
\ee
which in the $w=-1$ limit gives $\rho^{\Lambda_E}_E(z)=const\propto H^2_{E,0}\Omega_{\Lambda}$, due to the cancellation of the $\Omega$ factors. This is consistent with Equation~(\ref{MFELE}) and is expected from the fact that the cosmological constant part of the Lagrangian in Equation~(\ref{LLE}) is the same as in general~relativity.
}

At low redshift the effects of the cosmological constant are negligible, so that observationally it may not be possible to  distinguish between Equations~(\ref{MFELJ}) and~(\ref{MFELE}),  but~at at higher redshift the difference can become important. We leave to a future work the comparison with data to determine which dark energy model is in better agreement with high redshift observational data.



\begin{adjustwidth}{-\extralength}{0cm}

\reftitle{References}



\PublishersNote{}
\end{adjustwidth}

\begin{thebibliography}{999}
\makeatletter
\relax
\def\mn@urlcharsother{\let\do\@makeother \do\$\do\&\do\#\do\^\do\_\do\%\do\~}
\def\mn@doi{\begingroup\mn@urlcharsother \@ifnextchar [ {\mn@doi@}
  {\mn@doi@[]}}
\def\mn@doi@[#1]#2{\def\@tempa{#1}\ifx\@tempa\@empty \href
  {http://dx.doi.org/#2} {doi:#2}\else \href {http://dx.doi.org/#2} {#1}\fi
  \endgroup}
\def\mn@eprint#1#2{\mn@eprint@#1:#2::\@nil}
\def\mn@eprint@arXiv#1{\href {http://arxiv.org/abs/#1} {{\tt arXiv:#1}}}
\def\mn@eprint@dblp#1{\href {http://dblp.uni-trier.de/rec/bibtex/#1.xml}
  {dblp:#1}}
\def\mn@eprint@#1:#2:#3:#4\@nil{\def\@tempa {#1}\def\@tempb {#2}\def\@tempc
  {#3}\ifx \@tempc \@empty \let \@tempc \@tempb \let \@tempb \@tempa \fi \ifx
  \@tempb \@empty \def\@tempb {arXiv}\fi \@ifundefined
  {mn@eprint@\@tempb}{\@tempb:\@tempc}{\expandafter \expandafter \csname
  mn@eprint@\@tempb\endcsname \expandafter{\@tempc}}}



\bibitem[\protect\citeauthoryear{Riess et~al.}{Riess
  et~al.}{2022}]{Riess:2021jrx}
{Riess, A.G.;} 
 Yuan, W.; Macri, L.M.; Scolnic, D.; Brout, D.; Casertano, S.; Jones, D.O.; Murakami, Y.; An ; G.S.; Breuval, L.; Brink, T.G. A Comprehensive Measurement of the Local Value of the Hubble Constant with 1 km s\textsuperscript{$-$1} Mpc\textsuperscript{$-$1} Uncertainty from the Hubble Space Telescope and the SH0ES Team. \emph{Astrophys. J. Lett.} \textbf{2022}, \emph{934}, L7.
  \url{https://doi.org/10.3847/2041-8213/ac5c5b}.

\bibitem[\protect\citeauthoryear{Aghanim et~al.}{Aghanim
  et~al.}{2020}]{Planck:2018vyg}
Aghanim, N. et al. [Planck Collaboration] \emph{Planck} 2018 results. \emph{Astron. Astrophys.} \textbf{2020}, \emph{641}, {A6.} 
  \url{https://doi.org/10.1051/0004-6361/201833910}.


\bibitem{Freedman:2024eph}
Freedman, W.L.; Madore, B.F.; Hoyt, T.J.; Jang, I.S.; Lee, A.J.; Owens, K.A.
Status Report on the Chicago-Carnegie Hubble Program (CCHP): Measurement of the Hubble Constant Using the Hubble and James Webb Space Telescopes.
\emph{Astrophys. J.} \textbf{2025}, \emph{985}, 203.
\url{https://doi.org/10.3847/1538-4357/adce78}.

\bibitem{Jiang:2024tll}
Jiang, J.Q.; Liu, W.; Zhan, H.; Hu, B.
Explanation of high redshift luminous galaxies from JWST by an early dark energy model.
\emph{Phys. Rev. D} \textbf{2025}, \emph{111}, 023519.
\url{https://doi.org/10.1103/PhysRevD.111.023519}.


\bibitem{Liu:2024yan}
Liu, W.; Zhan, H.; Gong Y.; Wang, X.
Can early dark energy be probed by the high-redshift galaxy abundance?
\emph{Mon. Not. Roy. Astron. Soc.} \textbf{2024}, \emph{533}, 860--871;
Erratum in \emph{Mon. Not. Roy. Astron. Soc.} \textbf{2025}, \emph{538}, 1863.
\url{https://doi.org/10.1093/mnras/stae1871}.

\bibitem{Romano:2016utn}
Romano, A.E. Hubble trouble or Hubble bubble? \emph{Int. J. Mod. Phys. D} \textbf{2018}, \emph{27}, 1850102.
\url{https://doi.org/10.1142/S021827181850102X}.


\bibitem[\protect\citeauthoryear{Aluri et~al.}{Aluri
  et~al.}{2023}]{Aluri:2022hzs}
Aluri, P.K.; Cea, P.; Chingangbam, P.; Chu, M.C.; Clowes, R.G.; Hutsemékers, D.; Kochappan, J.P.; Lopez, A.M.; Liu, L.; Martens, N.C.; et~al. Is the observable Universe consistent with the cosmological principle? \emph{Class. Quant. Grav.} \textbf{2023}, \emph{40}, 094001.
  \url{https://doi.org/10.1088/1361-6382/acbefc}.

\bibitem[\protect\citeauthoryear{Di~Valentino et~al.,}{Di~Valentino
  et~al.}{2021}]{DiValentino:2021izs}
Di Valentino, E.; Mena, O.; Pan, S.; Visinelli, L.; Yang, W.; Melchiorri, A.; Mota, D.F.; Riess, A.G.; Silk, J. In the realm of the Hubble tension---A review of solutions. \emph{Class. Quant. Grav.} \textbf{2021}, \emph{38}, 153001.
  \url{https://doi.org/10.1088/1361-6382/ac086d}.

\bibitem[\protect\citeauthoryear{Montani, De~Angelis, Bombacigno  \&
  Carlevaro}{Montani et~al.}{2023}]{Montani:2023xpd}
Montani, G.; DeAngelis, M.; Bombacigno, F.; Carlevaro, N. Metric \emph{f}(\emph{R}) gravity with dynamical dark energy as a scenario for the Hubble tension. \emph{Mon. Not. Roy. Astron. Soc.} \textbf{2023}, \emph{527}, L156--L161. \url{https://doi.org/10.1093/mnrasl/slad159}.

\bibitem[\protect\citeauthoryear{Gleyzes, Langlois, Piazza  \&
  Vernizzi}{Gleyzes et~al.}{2013}]{Gleyzes:2013ooa}
Gleyzes, J.; Langlois, D.; Piazza, F.; Vernizzi, F. Essential building blocks of dark energy. \emph{J. Cosmol. Astropart. Phys.} \textbf{2013}, \emph{8}, 25.
  \url{https://doi.org/10.1088/1475-7516/2013/08/025}.

\bibitem[\protect\citeauthoryear{C\^ot\'e, Faraoni  \& Giusti}{C\^ot\'e
  et~al.}{2019}]{Cote:2019kbg}
C\^ot\'e, J.; Faraoni, V.; Giusti, A. Revisiting the conformal invariance of Maxwell's equations in curved spacetime. \emph{Gen. Rel. Grav.} \textbf{2019}, \emph{51}, 117.
  \url{https://doi.org/10.1007/s10714-019-2599-x}.

\bibitem[\protect\citeauthoryear{D'Inverno}{D'Inverno}{1992}]{inverno:1992}
D'Inverno R. \emph{Introducing Einstein's Relativity}; Clarendon Press: {Oxford, UK,} 
 1992.

\bibitem{Dabrowski:2008kx}
M.~P.~Dabrowski, J.~Garecki and D.~B.~Blaschke,
Conformal transformations and conformal invariance in gravitation.
\emph{Annalen Phys.} \textbf{2009}, \emph{18}, 13--32.
\url{https://doi.org/10.1002/andp.200810331}.

\bibitem[\protect\citeauthoryear{Uzan, Pernot-Borr\`as  \& Berg\'e}{Uzan
  et~al.}{2020}]{Uzan:2020aig}
Uzan, J.-P.; Pernot-Borr\`as, M.; Berg\'e, J. Effects of a scalar fifth force on the dynamics of a charged particle as a new experimental design to test chameleon theories. \emph{Phys. Rev. D} \textbf{2020}, \emph{102}, 044059. 
  \url{https://doi.org/10.1103/PhysRevD.102.044059}.

\makeatother

\bibitem[\protect\citeauthoryear{Romano \& Sakellariadou}{Romano \&
  Sakellariadou}{2023}]{Romano:2023ozy}
Romano, A.E.; Sakellariadou, M. Mirage of Luminal Modified Gravitational-Wave Propagation. \emph{Phys. Rev. Lett.} \textbf{2023}, \emph{130}, 231401. \url{https://doi.org/10.1103/PhysRevLett.130.231401}.

\bibitem[\protect\citeauthoryear{Chiba \& Yamaguchi}{Chiba \&
  Yamaguchi}{2013}]{Chiba:2013mha}
Chiba, T.; Yamaguchi, M. Conformal-frame (in)dependence of cosmological observations in scalar-tensor theory. \emph{J. Cosmol. Astropart. Phys.} \textbf{2013}, \emph{10}, 40. \url{https://doi.org/10.1088/1475-7516/2013/10/040}.


\bibitem[\protect\citeauthoryear{Deruelle \& Sasaki}{Deruelle \&
  Sasaki}{2011}]{Deruelle:2010ht}
Deruelle, N.; Sasaki, M. Conformal Equivalence in Classical Gravity: The Example of ``Veiled'' General Relativity. In \emph{{Cosmology, Quantum Vacuum and Zeta Functions}}; Springer Proceedings in Physics; {Springer: Berlin/Heidelberg, Germany,} 2011; Volume 137, pp.~247--260. 
  \url{https://doi.org/10.1007/978-3-642-19760-4\_23}.

\bibitem[\protect\citeauthoryear{Rondeau \& Li}{Rondeau \&
  Li}{2017}]{Rondeau:2017xck}
Rondeau, F.; Li, B. Equivalence of cosmological observables in conformally related scalar tensor theories. \emph{Phys. Rev. D} \textbf{2017}, \emph{96}, 124009. \url{https://doi.org/10.1103/PhysRevD.96.124009}.

\bibitem[\protect\citeauthoryear{Paraskevas \& Perivolaropoulos}{Paraskevas \&
  Perivolaropoulos}{2023}]{Paraskevas:2023aae}
Paraskevas, E.A.; Perivolaropoulos, L. Effects of a Late Gravitational Transition on Gravitational Waves and Anticipated Constraints. \emph{Universe} \textbf{2023}, \emph{9}, 317.
  \url{https://doi.org/10.3390/universe9070317}.

\bibitem[\protect\citeauthoryear{Marra \& Perivolaropoulos}{Marra \&
  Perivolaropoulos}{2021}]{Marra:2021fvf}
Marra V.; Perivolaropoulos L. Rapid transition of \emph{G}\textsubscript{eff} at \emph{z}\textsubscript{\emph{t}} $\simeq$ 0.01 as a possible solution of the Hubble and growth tensions. \emph{Phys. Rev. D} \textbf{2021}, \emph{104}, L021303. \url{https://doi.org/10.1103/PhysRevD.104.L021303}.


\bibitem[\protect\citeauthoryear{Schiavone, Montani  \& Bombacigno}{Schiavone
  et~al.}{2023}]{Schiavone:2022wvq}
Schiavone, T.; Montani, G.; Bombacigno, F. \emph{f}(\emph{R}) gravity in the Jordan frame as a paradigm for the Hubble tension. \emph{Mon. Not. Roy. Astron. Soc.} \textbf{2023}, \emph{522}, L72--L77. \url{https://doi.org/10.1093/mnrasl/slad041}.

\bibitem[\protect\citeauthoryear{Mazo, Romano  \& Quintero}{Mazo
  et~al.}{2022}]{Mazo:2022auo}
Mazo, B.Y.D.V.; Romano, A.E.; Quintero, M.A.C. H\textsubscript{0} tension or \emph{M} overestimation? \emph{Eur. Phys. J. C} \textbf{2022}, \emph{82}, 610. \url{https://doi.org/10.1140/epjc/s10052-022-10526-3}.

\bibitem[\protect\citeauthoryear{Scolnic et~al.,}{Scolnic
  et~al.}{2018}]{Scolnic_2018}
Scolnic, D.M.; Jones, D.O.; Rest, A.; Pan, Y.C.; Chornock, R.; Foley, R.J.; Huber, M.E.; Kessler, R.; Narayan, G.; Riess, A.G.; et~al. The Complete Light-curve Sample of Spectroscopically Confirmed SNe Ia from Pan-STARRS1 and Cosmological Constraints from the Combined Pantheon Sample. \emph{Astrophys. J.} \textbf{2018}, \emph{859}, 101. 
  \url{https://doi.org/10.3847/1538-4357/aab9bb}.

\bibitem{Linder:2015rcz}
Linder, E.V.; Seng{\"o}r, G.; Watson, S.
Is the Effective Field Theory of Dark Energy Effective?
\emph{J. Cosmol. Astropart. Phys.} \textbf{2016}, \emph{5}, 53.
\url{https://doi.org/10.1088/1475-7516/2016/05/053}.

\bibitem{Ishak:2024jhs}
Ishak, M.; Pan, J.; Calderon, R.; Lodha, K.; Valogiannis, G.; Aviles, A.; Niz, G.; Yi, L.; Zheng, C.; Garcia-Quintero, C.; et~al.
Modified Gravity Constraints from the Full Shape Modeling of Clustering Measurements from DESI 2024. \emph{arXiv} \textbf{2024}, arXiv:2411.12026.

\bibitem[\protect\citeauthoryear{Efstathiou}{Efstathiou}{2021}]{Efstathiou:2021ocp}
Efstathiou G. To H\textsubscript{0} or not to H\textsubscript{0}? \emph{Mon. Not. Roy. Astron. Soc.} \textbf{2021}, \emph{505}, 3866--3872. \url{https://doi.org/10.1093/mnras/stab1588}.



\makeatother
\end{thebibliography}
\end{document}